**Personalized Danger Signals Reprocessing: New Online Group Intervention for Chronic Pain**

Carmit Himmelblau Gat, Natalia Polyviannaya, Julia Zatulovsky, Pavel Goldstein

## Abstract

Background. Chronic pain affects roughly one in five adults, much of it chronic primary, or nociplastic, pain. Psychological therapy reaches few who need it, and remote programmes produce small, poorly maintained effects. Personalized Danger Signals Reprocessing (PDSR) is an eight-week coached online group intervention built on predictive processing, acting at once on the danger prior, on the actions that confirm it, and on the precision assigned to bodily signals.

Methods. Of 47 individuals screened with chronic non-cancerous, non-neuropathic pain, 39 women (83.0%) were randomized to PDSR (n = 19) or a waiting list (n = 20). Mean age was 40.7 years (SD 7.1). Outcomes were collected at baseline, week 4 and week 8, and pain intensity again at six months in both arms, and analysed with linear mixed-effects models. Cohen's d is standardised on the total standard deviation.

Results. At week 8 the PDSR group reported lower pain intensity than the waiting list (d = -1.43, 95% CI [-2.10, -0.77]), lower pain interference (d = -1.78 [-2.45, -1.10]) and higher well-being (d = 1.55 [0.89, 2.22]), all $p < .001$; the arms did not differ in pain intensity at week 4 (d = -0.50, p = .119). Catastrophizing (d = -1.61), anxiety (d = -1.59) and insomnia (d = -1.81) were also lower at week 8; for depression the time-by-group interaction was not significant (p = .151). Fifteen of 19 PDSR participants (79%) achieved at least a 30% reduction in pain and 12 (63%) at least 50%, against 7 (35%) and 3 (15%) on the waiting list. The within-arm gain held at six months: more then 50% of the PDSR group demonstrated complete or near-complete recovery. In an exploratory mediation, catastrophizing at week 4 carried the effect on week-8 pain (indirect effect -1.63 [-2.72, -0.72]).

Conclusions. PDSR produced large reductions in pain, interference and psychological comorbidity relative to a waiting list. The trial is small, single-sex, unblinded and waiting-list controlled, so expectancy cannot be separated from the proposed mechanism; an active-controlled trial is the next step.

## Introduction

Chronic pain affects roughly one in five adults and is a leading cause of disability (Rikard et al., 2023). Much of it is chronic primary pain: persisting beyond three months, with significant emotional distress or functional disability, not better accounted for by another condition (Treede et al., 2019; Nicholas et al., 2019), and mechanistically nociplastic (Kosek et al., 2021).
This burden falls disproportionately on women (Mogil, 2020). Its size depends on case definition - referred fibromyalgia samples are over 90% female, population samples nearer 60% (Wolfe et al., 2018) - but the direction is consistent, and women's pain is more often disbelieved and under-treated (Samulowitz et al., 2018). We therefore enrolled women only: they carry most of this burden and most of the unmet need, and at this sample size a single-sex trial removes variance we could not model.
A new generation of psychological treatments has raised what is achievable: pain reprocessing therapy left two thirds of patients with chronic back pain free or nearly free of pain against a fifth on placebo, held at one year (Ashar et al., 2022), far beyond conventional cognitive-behavioural therapy (Williams et al., 2020). Reaching people is the problem. Only 3.8% of US adults with chronic pain report any psychological treatment (Groenewald et al., 2022), constrained by cost, availability and too few trained clinicians (Schatman & Fortino, 2020). Remote delivery addresses patient-side reach (Rosser et al., 2023); group delivery addresses provider-side capacity, where services cannot meet demand (Choiniere et al., 2010) and one clinician treats a cohort at outcomes comparable to individual therapy (Turner-Stokes et al., 2003). McDermott and Keefe (2026) name both gaps. Yet remote programmes reach far and change little: improvements are small and rarely maintained, in trials largely self-guided (Eccleston et al., 2014).
That appraisal finds these treatments so overlapping that no unique active ingredient is identifiable; predictive processing offers a frame for specifying one. Pain is inferred, not read off the body: the brain combines prior expectations with ascending evidence weighted by its estimated precision (Mahajan & Seymour, 2025). Persistent pain is then an over-precise danger prior the system defends rather than updates (Kube et al., 2020): expectation biases the learning that would correct it (Jepma et al., 2018), and each protective act supplies the evidence keeping it in place (Milde et al., 2023). Revising a deeply connected prior disturbs a web of adjacent predictions, so defending it is cheaper.

Personalized Danger Signals Reprocessing (PDSR) is, to our knowledge, the first framework built wholly on predictive coding to act on all three constituents of the model at once. On the prior, psychoeducation reconceptualises the symptom as neuroplastic and reversible (Moseley & Butler, 2015), nocebo expectancies are corrected and positive expectancy cultivated. On active inference, participants learn to recognise the states corroborating the prior and substitute states incompatible with threat - an Investigator stance of curiosity, an Observer stance of non-judgmental attention - while graded exposure corrects fear-avoidance (Vlaeyen & Linton, 2000). On precision, somatic calibration - the novel component - trains fine interoceptive and proprioceptive discrimination through unfamiliar movement, symmetrical-limb comparison and imagery, performed slowly and under her control; somatic tracking, applied in that therapy to the pain state, extends here to sensations aversive and pleasant; and mindfulness weights interoceptive evidence more heavily (Zeidan & Vago, 2016). The intent is to saturate a system locked in the danger prior with safety signals at every level, until the prediction error makes defending the prior unfavourable. Progress is indexed by control, positive expectancy and felt safety, not by pain.

We compared eight weeks of coached online group PDSR with a waiting-list control in women with chronic non-cancerous, non-neuropathic pain, hypothesising reduced pain intensity and interference, improved well-being, and parallel change in catastrophizing, anxiety, depression and sleep, with catastrophizing as the pre-specified mediator, tested as exploratory.

## Methods

### Participants

Forty-seven individuals with chronic non-cancerous, non-neuropathic pain responded to recruitment advertisements placed on social media platforms; 39 (83.0%) met the eligibility criteria and were enrolled. Eligibility required pain of at least three months' duration prior to enrollment. The trial was designed for women, in whom chronic primary pain conditions are substantially more prevalent, and recruitment was directed accordingly. Participants had a mean age of 40.74 years (SD = 7.10), a mean pain duration of 4.21 years (SD = 2.34), a mean baseline pain intensity of 5.49 (SD = 1.96), and a mean of 15.51 years of education (SD = 2.84). Participants were randomly assigned to the 8-week Personalized Danger Signals Reprocessing (PDSR) intervention (n = 19) or to a waiting-list control condition (n = 20). All participants provided informed consent. The study was approved by [PI: SUPPLY - ethics committee and approval number] and was registered at [PI: SUPPLY - registry and trial ID].

### Study design

The PDSR intervention was evaluated in a two-arm, parallel-group randomized controlled trial with a waiting-list comparator. Assessments were administered at baseline, at the midpoint (week 4), immediately post-intervention (week 8), and at a six-month follow-up.

The PDSR intervention was delivered online over 8 weeks, with weekly two-hour group sessions complemented by intensive chat-based support and guided self-practice exercises. The intervention integrated psychoeducation about chronic pain mechanisms, coaching techniques to enhance motivation and self-efficacy, mindfulness practices, and somatosensory exercises for interoceptive awareness and the modulation of pain perception (see Appendix 1).

Primary outcomes were pain intensity, pain interference, and well-being. Secondary outcomes were pain catastrophizing, anxiety, depression, and insomnia. All outcomes were collected at baseline, week 4, and week 8. Pain intensity alone was additionally collected at six months, in both arms. Participants allocated to the control group received general psychoeducation about chronic pain.

**Research Tools:**

1. **Pain intensity** was evaluated using the Numeric Rating Scale for Pain (NRS Pain), which has been validated for pain (Hawker et al., 2011; Thong et al., 2018).
2. **Pain interference** was evaluated using the Pain Interference Subscale of the Brief Pain Inventory-Short Form (BPI-Short). This subscale has demonstrated excellent internal consistency, with a Cronbach's alpha between 0.84 and 0.94, indicating a high level of reliability in capturing the impact of pain on an individual's daily activities (Gjeilo et al., 2007) .
3. **Well-being** was assessed using the World Health Organization-Five (WHO-5) Well-Being Index. The WHO-5 Well-Being Index is a globally recognized self-report measure that evaluates subjective psychological well-being. Comprising five items, this index asks respondents to rate aspects of their well-being over the past two weeks, including mood, vitality, and general interests. Responses are scored on a 6-point scale from 0 (at no time) to 5 (all of the time), with higher scores indicating greater well-being. The WHO-5 Well-Being Index is noted for its brevity, ease of use, and strong psychometric properties. It has been validated across a wide range of populations and settings, proving to be a sensitive and specific tool for detecting states of depression and anxiety. It exhibits high internal consistency, with a Cronbach's alpha typically exceeding 0.84, affirming its reliability as a comprehensive gauge of individual well-being (Sischka et al., 2020).

4. **Anxiety** levels were measured employing the General Anxiety Disorder 7-item scale (GAD-7). The GAD-7 questionnaire is recognized for its validity and reliability in both clinical settings and research contexts. It effectively quantifies the severity of general anxiety symptoms, offering a succinct yet comprehensive overview of an individual's anxiety status. The scale has demonstrated robust psychometric properties, including a high degree of internal consistency, with a Cronbach's alpha coefficient of 0.92, underscoring its reliability for assessing general anxiety disorders (Spitzer et al., 2006).
5. **Depression** levels were assessed using the Patient Health Questionnaire-9 (PHQ-9). The PHQ-9 is a widely utilized and validated instrument designed to screen for the presence and severity of depressive symptoms. The questionnaire is esteemed for its clinical and research utility, providing a reliable measure of depression severity over the previous two weeks. It has demonstrated internal consistency and reliability, with a Cronbach's alpha of 0.83, affirming its effectiveness as a diagnostic and evaluative tool for depression (Costantini et al., 2021).
6.
7. **Insomnia symptoms were evaluated with the Pittsburgh Insomnia Rating Scale (PIRS-20), a 20-item self-report measure of insomnia severity rated 0-3, yielding a total score from 0 to 60, with higher scores indicating greater insomnia. Nineteen of the 20 items were administered; total scores were prorated as (sum of the 19 answered items / 19) x 20, per the instrument's scoring rule for a single missing item. The PIRS-20 has demonstrated good internal consistency and discriminates primary insomnia from good sleepers (Moul et al., 2002).**
8. **Pain catastrophizing** was evaluated with Pain catastrophizing scale (PCS) which is a 13-item questionnaire that quantifies an individual's tendency to focus on and magnify pain sensations and to feel helpless in the face of pain, as well as the degree to which they ruminate about their pain. Respondents rate statements based on their experiences on a 5-point scale, ranging from 0 (not at all) to 4 (all the time), with higher scores indicating a greater level of pain catastrophizing. The PCS has demonstrated internal consistency, with a Cronbach's alpha ranging from 0.87 to 0.95 across various studies, validating its efficacy as a reliable measure of pain catastrophizing (Leung, 2012).

**Data analysis**

All outcomes were analyzed with linear mixed-effects models fitted in R 4.3.3 (R Core Team, 2024) using lme4 and lmerTest, modelling each outcome as a function of time point, group, and their interaction. Models included subject-specific random intercepts. Random slopes were not estimable in this design: with 39 participants measured at three time points, a time-by-subject random-slope term requires as many random effects as there are observations (117), and the model is unidentifiable. Degrees of freedom were approximated by the Kenward-Roger method for the between-group contrasts and by the Satterthwaite method for the omnibus F tests. Between-group differences at each time point were estimated with estimated marginal means and Tukey-adjusted pairwise contrasts. Effect sizes are reported as Cohen's d standardised on the total standard deviation. Tukey adjustment was applied for post-hoc analysis.. Baseline characteristics are compared descriptively using standardised differences rather than significance tests, following CONSORT 2025. Internal consistency was estimated in the present sample using Cronbach's alpha with bootstrap confidence intervals. Responder analyses used the conventional 30% and 50% reduction thresholds in pain intensity. A half-longitudinal mediation model tested whether pain catastrophizing at week 4 carried the effect of group allocation on pain intensity at week 8, with both the mediator and the outcome adjusted for their baseline values; confidence intervals for the indirect effect were obtained by the distribution-of-the-product method (Tofighi & MacKinnon, 2011) and corroborated by Monte Carlo simulation.

## Results

### Sample Characteristics

Of 47 individuals screened, 39 (83.0%) met the eligibility criteria and were randomized, 19 to PDSR and 20 to the waiting list. [PI: SUPPLY - reasons for the 8 exclusions.] All 39 participants completed the baseline, week-4 and week-8 assessments and the six-month follow-up; there were no dropouts and no missing outcome data, so the randomized and analyzed samples are identical. Baseline characteristics were similar across arms (Table 1); standardised differences were 0.26 for age, 0.22 for pain duration, 0.04 for years of education, and 0.30 for baseline pain intensity. Participants had a mean age of 40.74 years (SD = 7.10), had lived with chronic pain for a mean of 4.21 years (SD = 2.34), and had completed a mean of 15.51 years of education (SD = 2.84).

Internal consistency in the present sample was acceptable to good for every multi-item scale: pain interference alpha = 0.81, 95% CI [0.70, 0.89]; pain catastrophizing 0.89 [0.83, 0.94]; anxiety 0.84 [0.75, 0.91]; depression 0.84 [0.75, 0.90]; well-being 0.77 [0.63, 0.87]; and insomnia 0.89 [0.83, 0.93]. All item-total correlations were positive, and no item required reverse keying.

**Table 1. Baseline Characteristics of Study Participants**

| | PDSR (N=19) | Waiting List (N=20) | Overall (N=39) |
|---|---|---|---|
| **Demographics** | | | |
| Age (years) | 41.68 (6.96) | 39.85 (7.30) | 40.74 (7.10) |
| Years with Chronic Pain | 4.47 (2.78) | 3.95 (1.88) | 4.21 (2.34) |
| Years of Education | 15.58 (2.73) | 15.45 (3.00) | 15.51 (2.84) |
| Baseline Pain Levels | 5.79 (1.99) | 5.20 (1.94) | 5.49 (1.96) |
| **Medication Profile** | | | |
| Current Analgesic Use, n (%) | 16 (84.2) | 17 (85.0) | 33 (84.6) |
| Number of Pain Medications† | 2.3 (1.4) | 2.1 (1.2) | 2.2 (1.3) |
| **Medication Classes, n (%)** | | | |
| NSAIDs (regular use) | 11 (57.9) | 12 (60.0) | 23 (59.0) |
| Acetaminophen (regular) | 9 (47.4) | 8 (40.0) | 17 (43.6) |
| Opioid analgesics | 5 (26.3) | 6 (30.0) | 11 (28.2) |
| Antidepressants for pain‡ | 8 (42.1) | 7 (35.0) | 15 (38.5) |
| Anticonvulsants for pain§ | 4 (21.1) | 5 (25.0) | 9 (23.1) |
| Topical analgesics | 7 (36.8) | 6 (30.0) | 13 (33.3) |
| **Primary pain diagnosis** | | | |
| Low-back pain (LBP) | 6 (31.6 %) | 8 (40.0 %) | 14 (35.9 %) |
| Migraine | 5 (26.3 %) | 4 (20.0 %) | 9 (23.1 %) |
| Irritable-bowel syndrome (IBS) | 2 (10.5 %) | 1 ( 5.0 %) | 3 ( 7.7 %) |
| Fibromyalgia (FM) | 4 (21.1 %) | 2 (10.0 %) | 6 (15.4 %) |
| Neck pain | 1 ( 5.3 %) | 2 (10.0 %) | 3 ( 7.7 %) |
| Knee pain | 1 ( 5.3 %) | 3 (15.0 %) | 4 (10.3 %) |

*Values are mean (SD) unless the row is marked n (%); the number of pain medications is mean (SD). Following CONSORT 2025, groups are compared descriptively rather than by significance test; standardised differences for the demographic variables are reported in the text. [PI: SUPPLY - definitions for the three footnote markers, which are undefined in the table.]*

**Primary outcomes**

**Pain Levels**

A mixed-effects model analysis revealed a significant interaction between time and group for pain intensity ($F(2, 74) = 9.44$, $p < 0.001$), indicating differing patterns of change across time between the PDSR and waiting list groups. At baseline, pain intensity was similar between the PDSR (M =

5.79, SE = 0.43) and waiting list groups (M = 5.20, SE = 0.42; M difference = 0.589, SE = 0.600; t(102) = 0.98, p = 0.328; Cohen's d = 0.31, 95% CI [-0.32, 0.95]). At the midpoint the difference favoured the PDSR group but did not reach significance (M difference = -0.942, SE = 0.600; t(102) = -1.57, p = 0.119; Cohen's d = -0.50, 95% CI [-1.14, 0.14]). By the post-intervention assessment the PDSR group reported substantially lower pain intensity (M = 2.32, SE = 0.43) than the waiting list group (M = 5.00, SE = 0.42; M difference = -2.684, SE = 0.600; t(102) = -4.48, p < 0.001; Cohen's d = -1.43, 95% CI [-2.10, -0.77]) (Figure 1A).

**Pain Interference**

A mixed-effects model analysis revealed a significant interaction between time and group for pain interference (F(2, 111) = 9.19, p < 0.001), indicating that change in pain interference over time differed between the PDSR and waiting list groups. At baseline the two groups were equivalent (PDSR M = 4.72, SE = 0.29; waiting list M = 4.72, SE = 0.28; M difference = 0.000, SE = 0.404; t(111) = 0.00, p = 0.999; Cohen's d = 0.00, 95% CI [-0.63, 0.64]). At the midpoint the groups did not differ (M difference = -0.266, SE = 0.404; t(111) = -0.66, p = 0.512; Cohen's d = -0.21, 95% CI [-0.85, 0.42]). By the post-intervention assessment the PDSR group reported markedly lower pain interference (M = 2.29, SE = 0.29) than the waiting list group (M = 4.53, SE = 0.28; M difference = -2.243, SE = 0.404; t(111) = -5.55, p < 0.001; Cohen's d = -1.78, 95% CI [-2.45, -1.10]) (Figure 1B).

**Well-Being**

A mixed-effects model analysis revealed a significant interaction between time and group for well-being (F(2, 74) = 10.70, p < 0.001), indicating differing patterns of change over time between the PDSR and waiting list groups. At baseline, well-being was comparable between the PDSR (M = 2.53, SE = 0.16) and waiting list groups (M = 2.69, SE = 0.15; M difference = -0.164, SE = 0.218; t(109.1) = -0.75, p = 0.454; Cohen's d = -0.24, 95% CI [-0.88, 0.40]). By the midpoint well-being in the PDSR group had increased relative to the waiting list group (M difference = 0.964, SE = 0.218; t(109.1) = 4.42, p < 0.001; Cohen's d = 1.42, 95% CI [0.75, 2.08]), and the advantage was maintained at the post-intervention assessment (PDSR M = 3.35, SE = 0.16; waiting list M = 2.29, SE = 0.15; M difference = 1.057, SE = 0.218; t(109.1) = 4.85, p < 0.001; Cohen's d = 1.55, 95% CI [0.89, 2.22]) (Figure 1C).

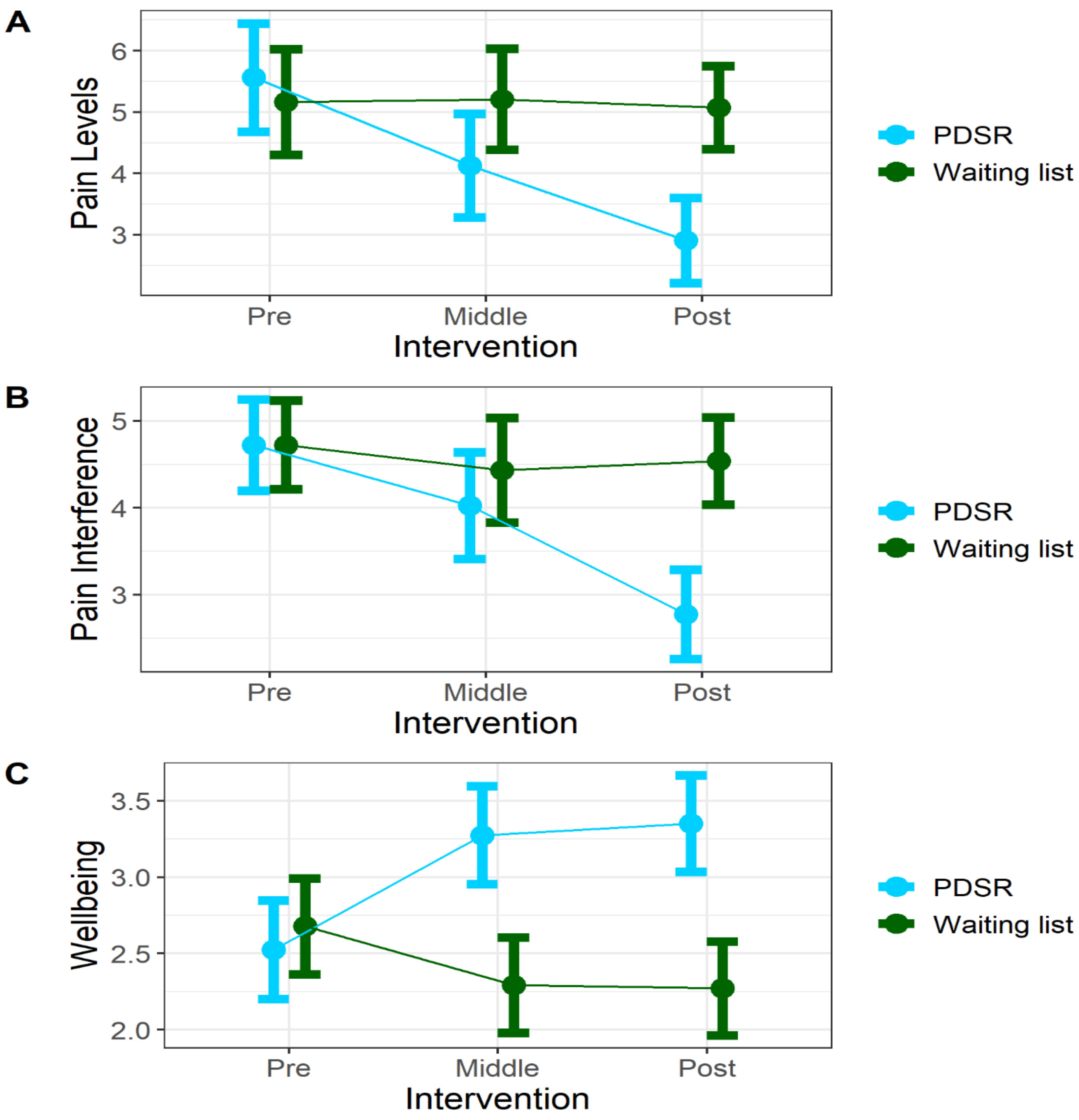

**Figure 1. Changes in primary outcomes across time for the PDSR and waiting list groups. (A) Pain intensity: pain intensity in the PDSR group did not differ significantly from the waiting list group at the midpoint, and was substantially lower by the post-intervention assessment.**

(B) Pain interference: The PDSR group showed significant reductions in pain interference over time, with large differences emerging by the post-intervention time point.

(C) Well-being: Well-being scores significantly improved in the PDSR group compared to the waiting list group, with large and sustained effects observed at both the midpoint and post-intervention time points. Error bars represent 95% confidence intervals.

**Secondary outcomes**

**Pain Catastrophizing**

A mixed-effects model analysis revealed a significant interaction between time and group for pain catastrophizing ($F(2, 74) = 7.36$, $p = 0.001$). At baseline the groups were comparable (PDSR M = 2.90, SE = 0.14; waiting list M = 3.05, SE = 0.14; M difference = -0.151, SE = 0.194; $t(108.8) = -0.78$, $p = 0.439$; Cohen's d = -0.25, 95% CI [-0.88, 0.39]). By the midpoint pain catastrophizing in the PDSR group had fallen well below the waiting list group (M difference = -1.054, SE = 0.194; $t(108.8) = -5.42$, $p < 0.001$; Cohen's d = -1.74, 95% CI [-2.41, -1.06]), and the difference persisted at the post-intervention assessment (PDSR M = 1.94, SE = 0.14; waiting list M = 2.91, SE = 0.14; M difference = -0.976, SE = 0.194; $t(108.8) = -5.02$, $p < 0.001$; Cohen's d = -1.61, 95% CI [-2.28, -0.94]). Notably, the reduction was already at its maximum by week 4 and did not deepen thereafter.

**Anxiety**

A mixed-effects model analysis revealed a significant interaction between time and group for anxiety ($F(2, 74) = 7.80$, $p < 0.001$). At baseline, anxiety was similar between the PDSR (M = 1.81, SE = 0.13) and waiting list groups (M = 1.73, SE = 0.12; M difference = 0.083, SE = 0.176; $t(110.6) = 0.47$, $p = 0.636$; Cohen's d = 0.15, 95% CI [-0.48, 0.79]). Anxiety in the PDSR group was lower than in the waiting list group at the midpoint (M difference = -0.495, SE = 0.176; $t(110.6) = -2.81$, $p = 0.006$; Cohen's d = -0.90, 95% CI [-1.55, -0.26]) and lower still at the post-intervention assessment (PDSR M = 0.57, SE = 0.13; waiting list M = 1.44, SE = 0.12; M difference = -0.871, SE = 0.176; $t(110.6) = -4.95$, $p < 0.001$; Cohen's d = -1.59, 95% CI [-2.26, -0.92]).

**Depression**

For depression, the interaction between time and group was not significant ($F(2, 111) = 1.92$, $p = 0.151$). Main effects were observed for time ($F(2, 111) = 8.47$, $p < 0.001$) and for group ($F(1, 111) = 5.15$, $p = 0.025$). At baseline the groups were equivalent (PDSR M = 1.67, SE = 0.14; waiting list M

= 1.65, SE = 0.13; M difference = 0.023, SE = 0.191; t(111) = 0.12, p = 0.907; Cohen's d = 0.04, 95% CI [-0.60, 0.67]), and they did not differ at the midpoint (M difference = -0.268, SE = 0.191; t(111) = -1.40, p = 0.165; Cohen's d = -0.45, 95% CI [-1.09, 0.19]). At the post-intervention assessment the PDSR group reported lower depression scores (M = 0.85, SE = 0.14) than the waiting list group (M = 1.36, SE = 0.13; M difference = -0.507, SE = 0.191; t(111) = -2.65, p = 0.009; Cohen's d = -0.85, 95% CI [-1.49, -0.20]). Because the omnibus interaction was not significant, this contrast should be read as exploratory.

**Insomnia**

A mixed-effects model analysis revealed a significant interaction between time and group for insomnia (F(2, 111) = 6.98, p = 0.001). At baseline, insomnia severity was comparable between the PDSR (M = 26.70, SE = 1.83) and waiting list groups (M = 28.37, SE = 1.78; M difference = -1.665, SE = 2.556; t(111) = -0.65, p = 0.516; Cohen's d = -0.21, 95% CI [-0.84, 0.43]). By the midpoint, insomnia severity in the PDSR group had fallen sharply relative to the waiting list group (M difference = -11.784, SE = 2.556; t(111) = -4.61, p < 0.001; Cohen's d = -1.48, 95% CI [-2.14, -0.81]), and the difference was larger still at the post-intervention assessment (PDSR M = 12.80, SE = 1.83; waiting list M = 27.26, SE = 1.78; M difference = -14.465, SE = 2.556; t(111) = -5.66, p < 0.001; Cohen's d = -1.81, 95% CI [-2.49, -1.13]).

**Clinical significance and responder rates**

At the post-intervention assessment, 15 of 19 participants in the PDSR group (79%) had achieved a reduction of at least 30% in pain intensity from baseline and 12 (63%) a reduction of at least 50%, compared with 7 of 20 (35%) and 3 of 20 (15%) in the waiting list group. Nine PDSR participants (47%) reported a pain intensity of 1 or less at week 8, and five (26%) reported no pain, compared with none in the waiting list group. At six months, 15 of 19 PDSR participants (79%) met the 50% reduction threshold, up from 12 (63%) at week 8, and 9 (47%) reported a pain intensity of 1 or less, compared with 4 of 20 (20%) and 1 of 20 (5%) in the waiting list group, which had by then received the recorded programme.

**Six-month follow-up**

In the PDSR group, mean pain intensity was 5.79 at baseline, 4.16 at week 4, 2.32 at week 8 and 2.05 at six months; the change from week 8 to six months was not significant (estimate = 0.263, SE

= 0.504, p = 0.954), indicating that the post-intervention gain was maintained. In the waiting list group, mean pain intensity was 5.20 at baseline, 5.10 at week 4, 5.00 at week 8 and 4.30 at six months (week 8 to six months: estimate = 0.700, SE = 0.491, p = 0.486). Because waiting-list participants were given access to the recorded intervention after the week-8 assessment, the six-month values in that arm do not represent an untreated course and no between-group comparison at six months is reported.

**Mediation by pain catastrophizing**

A half-longitudinal mediation model tested whether pain catastrophizing at week 4 carried the effect of group allocation on pain intensity at week 8, adjusting both the mediator and the outcome for their baseline values. Allocation to PDSR predicted lower pain catastrophizing at week 4 (a = -1.048, SE = 0.277), and pain catastrophizing at week 4 predicted pain intensity at week 8 (b = 1.554, SE = 0.260). The indirect effect was -1.628, 95% CI [-2.723, -0.715] by the distribution-of-the-product method and [-2.729, -0.712] by Monte Carlo simulation. The direct effect was no longer significant once the mediator was included (c′ = -0.968, p = 0.075). The reverse-ordered model, in which pain intensity at week 4 was tested as the mediator of the effect on pain catastrophizing at week 8, produced an indirect effect that did not differ from zero (-0.026, 95% CI [-0.142, 0.062]; Monte Carlo [-0.143, 0.063]), which is consistent with the hypothesised direction of influence. With 39 participants this mediation analysis is underpowered and is reported as exploratory; it establishes temporal ordering rather than causal mediation.

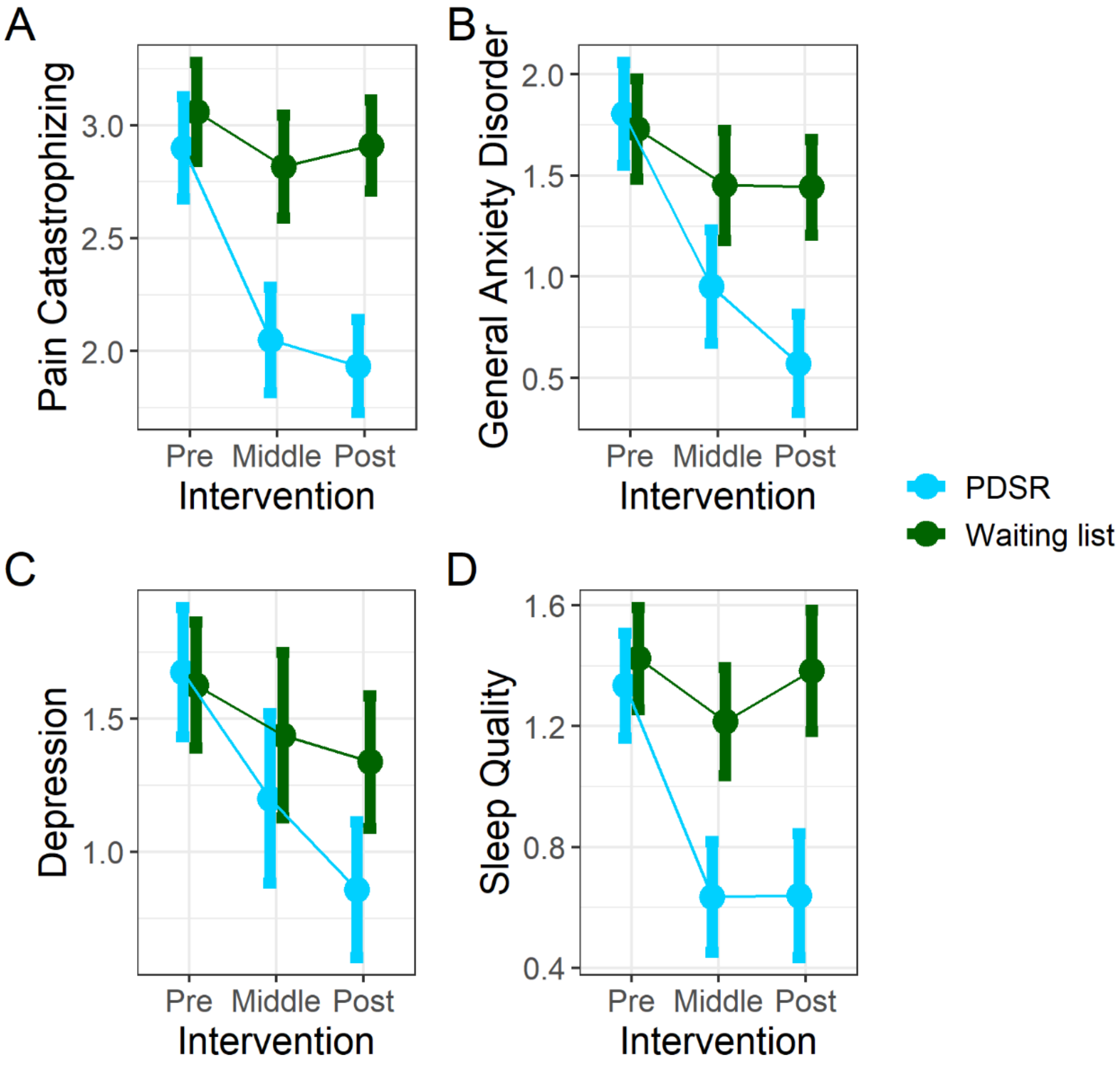

**Figure 2.** Changes in secondary outcomes across time for the PDSR and waiting list groups. (A) Pain catastrophizing: Pain catastrophizing decreased significantly in the PDSR group compared to the waiting list group, with large effect sizes observed at both the midpoint and post-intervention time points.

(B) Anxiety: Anxiety levels showed significant reductions in the PDSR group relative to the waiting list group, with moderate to large effect sizes observed at the midpoint and post-intervention time points.

(C) Depression: depression scores were lower in the PDSR group than in the waiting list group at the post-intervention assessment, although the time-by-group interaction was not significant and this difference is therefore exploratory.

(D) Insomnia: insomnia severity fell markedly in the PDSR group relative to the waiting list group, with large differences at both the midpoint and the post-intervention assessment. Error bars represent 95% confidence intervals.

**Discussion**

This study evaluates the initial efficacy of Personalized Danger Signals Reprocessing (PDSR), an online group intervention for chronic primary, or nociplastic, pain. Relative to a waiting-list control, the PDSR group improved on pain intensity, pain interference, well-being, pain catastrophizing, anxiety and insomnia, with effect sizes at week 8 ranging from d = 1.43 to d = 1.81 in absolute value. For depression the time-by-group interaction was not significant, and the week-8 difference should be read as exploratory. In the PDSR arm the reduction in pain intensity was maintained at six months.

**Interpreting the Findings:**

Our results further reinforce the importance of therapeutic strategies that target the mental processing of pain signals, consistent with the conceptual framework of nociplastic pain as driven by aberrant neural processing rather than solely tissue damage (Fitzcharles et al., 2021; Galli, 2023). The study's findings align with existing evidence supporting psychotherapeutic approaches such as emotional awareness and expression therapy (EAET) (Yarns et al., 2020) and pain reprocessing therapy (PRT) (Ashar et al., 2022) for chronic pain management.

Pain reprocessing therapy (PRT) aims to shift patients' beliefs about the causes and threat value of pain. It has demonstrated substantial and durable pain relief in chronic back pain patients (Ashar et al., 2022). A key mechanism underlying PRT's effectiveness appears to be the reattribution of pain to mind or brain processes rather than tissue damage. This reattribution was associated with pain relief and mediated the effects of PRT on long-term pain intensity (Ashar et al., 2023). Importantly, in PRT, the therapist actively guides the patient through the treatment process, playing a central role in facilitating cognitive and emotional shifts (Ashar et al., 2022; Ashar et al., 2023).

Emotional awareness and expression therapy (EAET) emphasizes emotional processing and central nervous system involvement in pain perception. It has achieved greater pain reduction than cognitive behavioral therapy (CBT) in older veterans with chronic musculoskeletal pain (Yarns et al., 2020). EAET focuses on helping patients identify and express suppressed emotions, which may contribute to pain persistence, thereby fostering emotional regulation and improved well-being.

Group PDSR belongs to the newer psychotherapeutic interventions that target the mental processing of pain signals. It draws on several existing traditions: like pain reprocessing therapy and emotional awareness and expression therapy it makes extensive use of pain neuroscience education and of the reprocessing of bodily sensations within a fear-avoidance framework, so that patients come to hold, rather than merely accept, the understanding that their pain does not signal tissue damage (Vergeld et al., 2021; Lumley & Schubiner, 2019a).

PDSR differs from these approaches in several respects. One is the role of the therapist: in PDSR participants take primary responsibility for their own learning, and the therapist directs and sustains the group process rather than guiding each patient individually.

PDSR interventions incorporate additional components designed to enhance self-awareness and self-regulation. These include techniques for identifying and interpreting personal body sensations, building a personal roadmap to recognize individual cues of danger while structured and group chat support. Patients also learn to identify relaxation methods that are most effective for them. The approach employs coaching strategies that encourage curiosity and ownership of the healing process, fostering a more active and engaged role in managing pain.

The improvements in insomnia and the reduction in anxiety are noteworthy, since disturbed sleep and anxiety are known to exacerbate chronic pain (Chen et al., 2022; Sheng et al., 2017; Tang et al., 2007). Pain catastrophizing also fell substantially. Depression scores were lower in the PDSR group at week 8, but the time-by-group interaction was not significant and this difference is therefore exploratory. Both catastrophizing and depression independently predict pain-related disability, with catastrophizing having the larger effect (Arnow et al., 2011), and chronic pain and depression are closely interrelated with catastrophizing playing a mediating role (Kardash et al., 2024; Kim et al., 2021; Glette et al., 2021). PDSR's integrative approach may therefore address comorbid conditions alongside pain itself, potentially interrupting the cycle in which each sustains the other.

**Potential Therapeutic Mechanisms:**

The therapeutic mechanisms at the core of PDSR involve a nuanced interplay of cognitive, emotional, and physiological processes. PDSR integrates a diverse array of approaches, including psychoeducation, coaching, mindfulness, and interoceptive reprocessing techniques. This comprehensive blend of methods forms the foundation of PDSR, which is delivered online as a form of teletherapy.

PDSR follows a structured curriculum built on danger-signal reprocessing, guiding participants to reframe their interpretation of painful sensations. In the predictive-processing terms set out in the Introduction, the intervention targets an over-precise danger prior and the weight the system assigns to bodily evidence. Those computational quantities are implemented in neural circuitry, and the same threat-processing network that the following paragraph describes is where a high-precision danger prior would be expected to leave its signature. The two levels of description are complementary rather than alternative: neither was measured here, and both are offered as an account of how PDSR might work rather than as a finding of this trial.

The amygdala is a central hub integrating emotional processing, pain modulation and threat appraisal. In chronic pain, altered connectivity between the amygdala and the periaqueductal gray, default mode network, salience network and central executive network is associated with heightened pain catastrophizing and emotional facilitation of pain [PI: SUPPLY - Timmers et al. (2021) and Mao et al. (2021) are cited here but neither appears in the reference list]. Elevated perceived threat enhances insula-prefrontal coupling, biasing perceptual decisions toward pain before a painful stimulus occurs (Johansson et al., 2024), and pain-related fear and catastrophizing are linked to disrupted amygdala-periaqueductal gray and amygdala-central executive connectivity (Meier et al., 2017; Jiang et al., 2016).
The therapeutic process facilitates a shift away from potentially catastrophic or threatening appraisals towards a more neutral and accepting perspective. Indeed, the significant reduction in pain catastrophizing observed in this study support previous findings (Petrini & Arendt-Nielsen, 2020) indicating that modifying maladaptive pain perceptions can lead to better pain outcomes. This aligns with PDSR's focus on reframing individuals' interpretations of pain signals, which may facilitate neural plasticity and pain reduction (Doan et al., 2015; Mansour et al., 2014; Woolf, 2011).

Complemented by mindfulness and interoceptive practices, as well as coaching techniques, this cognitive reframing not only aids in emotional and physiological regulation but also diminishes the distress associated with pain. By addressing anxiety, insomnia and overall well-being alongside pain itself, PDSR offers a broad approach to chronic pain management.

The psychoeducation component, serving as the foundational framework of PDSR, demonstrated the role of participants' awareness regarding the nature of pain (De et al., n.d.; Gómez-De-regil, 2021). This educational approach guides participants through a progressive journey beginning with a limited understanding and evolving into active exploration and self-reflection regarding their own pain experiences. By introducing key concepts such as neuroplasticity of the brain, the dynamic relationship between pain and emotions, catastrophizing, the expectation management and other relevant ideas form clear and aware mindset and shift to a paradigm of biopsychosocial awareness.

Coaching techniques can help increase mental resilience and empower individuals to translate theoretical knowledge into precise actions for effective pain management. A previous study highlights the potential of coaching interventions to enhance mental resilience and well-being across various populations. A school-based health coaching program for middle school students significantly improved youth resilience (Lee et al., 2020). In the mental health field, coaching is proposed as a strategy to empower clients and assist healthcare professionals in implementing the recovery approach (McSharry & O'Grady, 2020). These studies collectively demonstrate the effectiveness of coaching techniques in building mental resilience and translating knowledge into actionable strategies across different contexts. In our study, the use of coaching techniques may have played a key role in contributing to the reduction of pain and other related symptoms.

The structured format of the program allows participants to progressively acquire new insights and develop essential skills, with intervals between sessions offering valuable opportunities to process and apply the learned material effectively. However, it is essential to acknowledge that this process unfolds uniquely for each participant, emphasizing the individualized nature of pain reprocessing and management within the program. The group format of the intervention capitalizes on social support and shared experiences, bolstering its therapeutic impact as evidenced by research on group therapy in chronic pain management (Nøst et al., 2022). Within the PDSR group context, the provision of group support and guidance through chat facilitates the exchange of insights and cultivates a strong sense of community. This inclusive environment may enable participants to embark on their journey of self-discovery with a sense of security, contributing to the reduction of anxiety levels. Through active engagement, participants could feel acknowledged, supported, and empowered to share valuable advice and experiences, boosting the engagement of other participants.

In the PDSR arm the reduction in pain intensity was maintained at six months, with no significant change between week 8 and follow-up. This maintenance is uncontrolled: waiting-list participants received the recorded programme after the week-8 assessment and their pain also declined over the same interval, though not significantly (estimate = 0.700, $p = .486$), so no between-group comparison at six months is available. With that caveat, the pattern is consistent with the durability reported for other psychological treatments. Pain reprocessing therapy produced lasting reduction relative to placebo and usual care at one year (Ashar et al., 2022), and complementary biopsychosocial treatments including self-hypnosis and cognitive-behavioural therapy have shown benefits up to twelve months (Bicego et al., 2021).

PDSR introduces an approach that not only integrates seamlessly into the therapeutic landscape but also offers a personalized perspective, presenting a pragmatic and effective solution for reducing chronic pain and associated mental comorbidities.

The pattern of results carries implications for practice, though the trial was not designed to test them. An online group format is, in principle, more scalable and less costly per patient than individual face-to-face therapy, and the responder rates observed here suggest the approach is worth evaluating at scale. No cost, clinician-time, uptake or implementation outcome was collected in this study, however, so scalability and cost-effectiveness remain claims about the delivery model rather than findings. Given the burden of chronic pain on individuals and health systems (T. J. Smith & Hillner, 2019), an adequately powered implementation study is the natural next step.

**Limitations and Future Directions:**

Several limitations qualify these findings. The comparator was a waiting list, so allocation is confounded with expectancy, attention and demand characteristics; the intervention's proposed mechanism cannot be separated from those non-specific effects, and an active or attention-matched control is required to do so. Neither participants nor the interventionist could be blinded, and all outcomes were self-reported. The trial was not powered a priori; the sample of 39 is small, and the mediation analysis in particular is underpowered and reported as exploratory. Three co-primary outcomes were analysed without a formal multiplicity hierarchy, and no adjustment was made across the seven outcomes, so the secondary results are exploratory. Waiting-list participants were given the recorded programme after week 8, which removes the six-month comparison. A set of item-level values was corrected after the initial data file was compiled, as described in the Methods; the responder analyses at the 30% and 50% thresholds are unaffected by that correction, while the counts of participants reporting no pain are sensitive to it. Participants were all women, which was by design but bounds generalisation to women who seek out and enrol in a remotely delivered programme. Finally, durability beyond six months is unknown, and the components of the intervention were not dismantled, so which elements carry the effect remains untested.

By addressing these limitations, future research can refine and extend the current findings, ultimately enhancing the clinical applicability and impact of PDSR in pain management.

**Conclusion:**

In summary, our investigation into the PDSR online intervention demonstrates its potential as a transformative approach in the realm of chronic pain management. The positive outcomes observed in pain reduction and mental health comorbidities within the PDSR group align with the broader trend in recognizing the limitations of traditional biomedical approaches and the importance of addressing psychological components associated with chronic pain. Study contributes to the growing body of evidence supporting the integration of digital and psychological interventions in pain management strategies. Further research is warranted to optimize PDSR and elucidate its long-term impact on individuals living with chronic pain.

**Appendix 1**

**PDSR Intervention Overview**

The PDSR intervention comprised an 8-week online group program designed to provide participants with a comprehensive approach to chronic pain management. Led by experienced pain management educators, the program integrated psychoeducation, coaching techniques, and mindfulness tools. The primary objective was to enhance participants' understanding of chronic pain, its underlying mechanisms, and its effects on daily life. Additionally, the course aimed to equip participants with non-pharmacological pain management strategies. These strategies were not only educational but also practical, emphasizing the importance of self-regulation and mental well-being.

Participants engaged in sessions conducted via Zoom, allowing for real-time interaction and seamless delivery of the course material. Each session spanned approximately 1.5 to 2 hours, ensuring ample time for participants to delve deeply into the content. This extended duration facilitated comprehensive coverage of the material and provided opportunities for in-depth discussions and active participation.

Design of the curriculum included this range of critical topics:

1. Physiology of Pain: Participants delved into the fundamental understanding of pain, its awareness, and the statistics related to chronic pain. They explored how pain becomes chronic and the potential secondary benefits of perceiving pain.

2. Pain and Fear: Session emphasized the connection between pain and fear, the importance of controlling pain, and the rational interpretation of pain perception. Body-oriented techniques and strategies for shifting attention were introduced.

3. Neuroplasticity and Chronic Pain: The course delved into the concept of neuroplasticity and its role in chronic pain. Participants learned about chronic pain scenarios, central sensitization, and factors contributing to neuroplastic pain.

4. Placebo and Nocebo: Participants explored the placebo effect, its connection to the mind and body, and its role in pain management. The session also addressed the phenomenon of nocebo.

5. Pain as Feedback: This session focused on the concept of pain as feedback and introduced an algorithm for managing neuroplastic pain. Participants learned to identify and address the five attention traps in mindfulness and engage in basic mindfulness training.

6. Resource Conditions: Participants worked on developing resource conditions, managing emotions, and recognizing cognitive and bodily signals of safety. They practiced safety signals and attention-shifting techniques.

7. Somatic Tracking and Exposition Therapy: The course explored somatic tracking and the principles of exposition therapy. Participants learned to work with intense sensations, the relationship between pain and touch, and the concept of safe avoidance. They also discussed the refocusing of attention and different types of threat alerts.

8. Resistance and Habit Formation: The final session emphasized overcoming resistance to change and creating new habits for pain management. Participants developed a personal set of keys to living pain-free, highlighting the importance of individual responsibility.
   To create a supportive and inclusive learning environment, participants were given access to a private chat group on Telegram. This group chat served as a virtual meeting place where participants could connect, share their experiences, ask questions, and provide valuable feedback. It was a space where both trainers and fellow participants actively interacted, fostering a sense of community and support throughout 8 weeks of the course.

**Appendix 3**
**Primary Outcomes**

**Figure 1. Pain intensity, pain interference and well-being in the PDSR and waiting-list groups. Pain intensity is shown at baseline, week 4, week 8 and six months; interference and well-being at baseline, week 4 and week 8.**

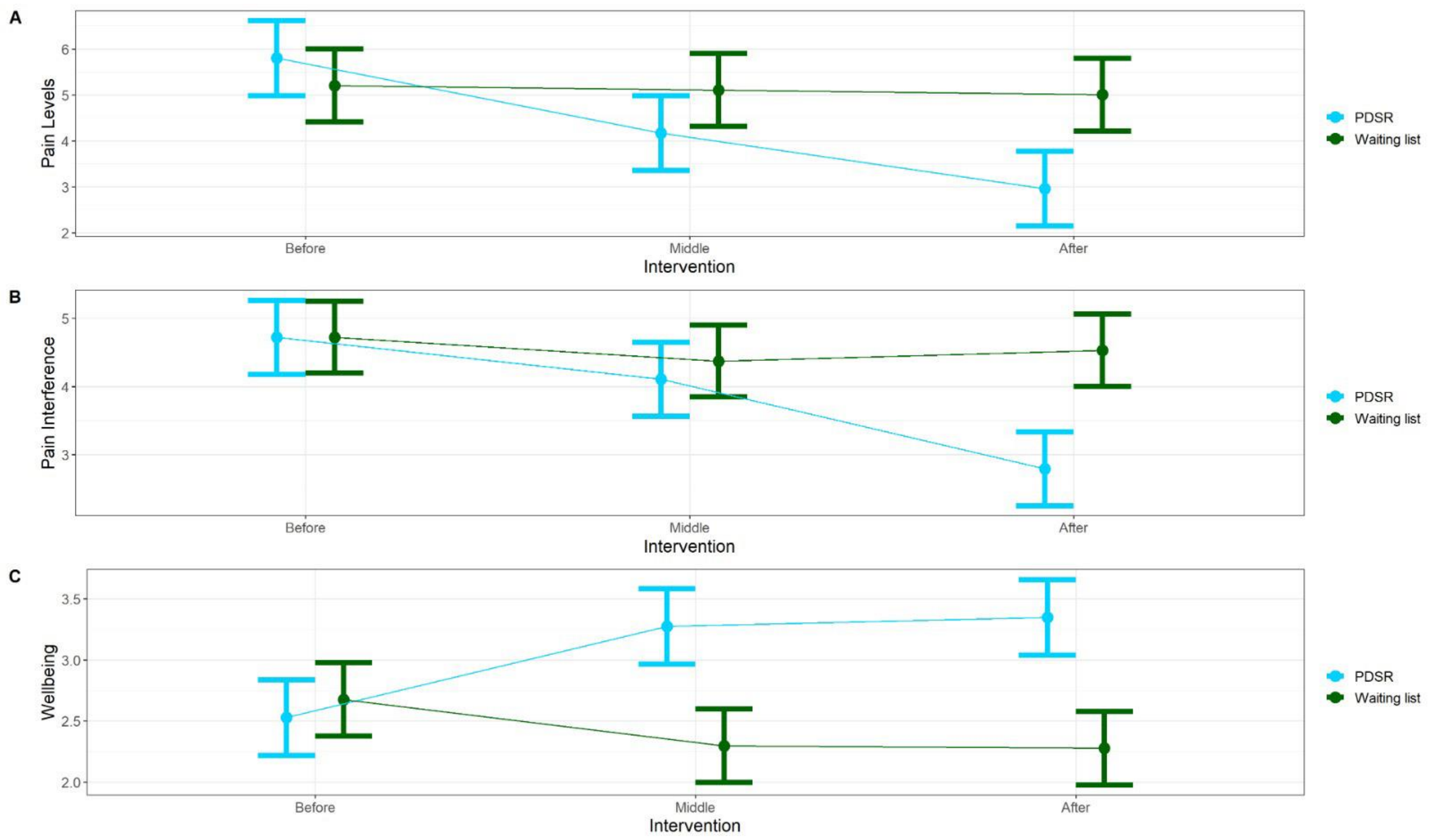

## Secondary Outcomes

**Figure 2. Pain catastrophizing, anxiety, depression and insomnia in the PDSR and waiting-list groups at baseline, week 4 and week 8.**

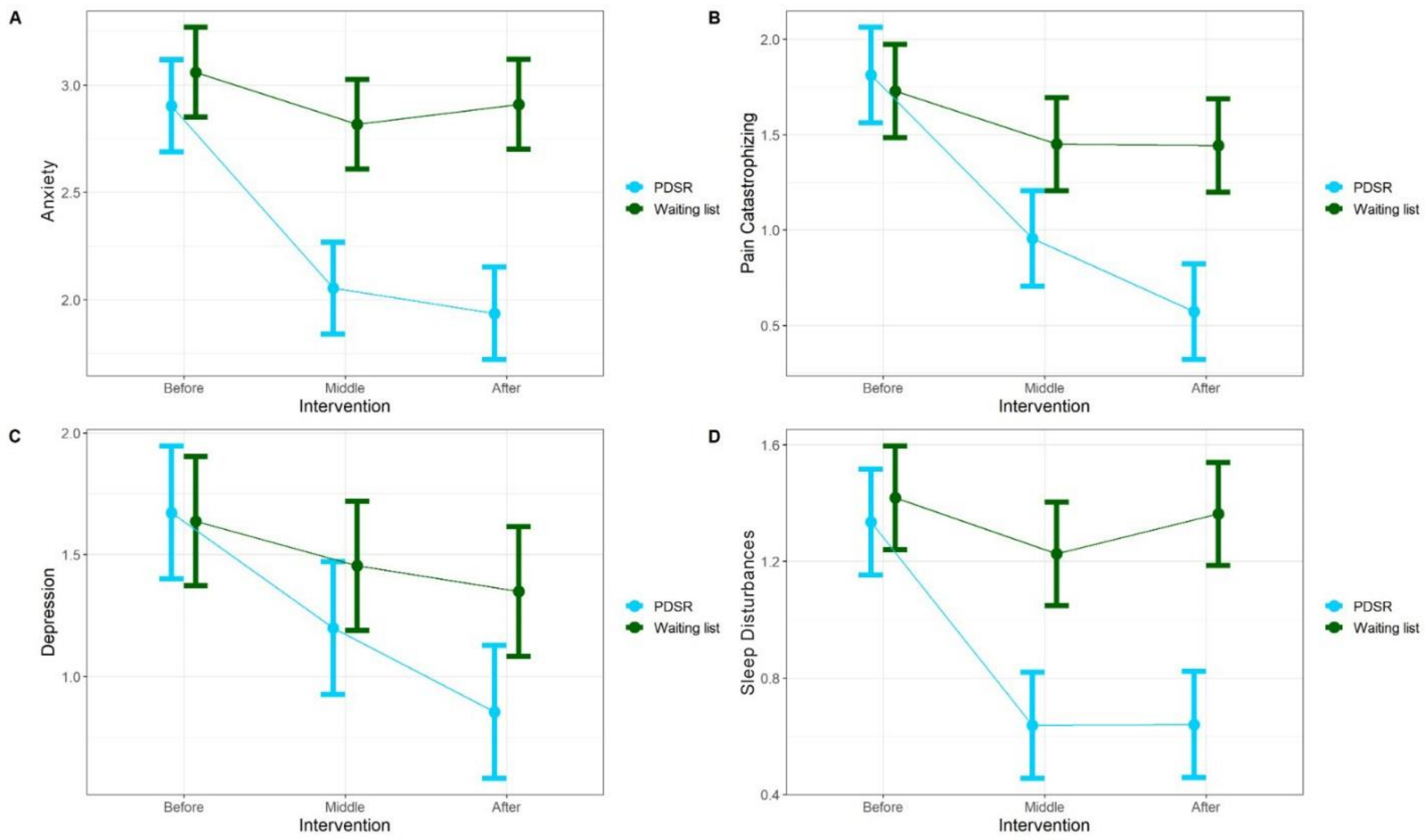
A
Anxiety
3.0
2.5
2.0
Before
Middle
After
Intervention
PDSR
Waiting list
B
Pain Catastrophizing
2.0
1.5
1.0
0.5
Before
Middle
After
Intervention
PDSR
Waiting list
C
Depression
2.0
1.5
1.0
Before
Middle
After
Intervention
PDSR
Waiting list
D
Sleep Disturbances
1.6
1.2
0.8
0.4
Before
Middle
After
Intervention
PDSR
Waiting list